\def\ln{\ell{n}}
\documentstyle[a4,12pt]{article}

\begin{document}
\begin{titlepage} \vspace{0.2in} 

\begin{center} {\LARGE \bf 
On the Covariance of the Mixmaster Chaoticity  
\\} 
\vspace*{2cm}

{\bf Giovanni Imponente  \\
Giovanni Montani  
}\\ 

\vspace*{2.8cm}
ICRA---International Center for Relativistic Astrophysics \\ 
Phys. Dept., University of Rome `La Sapienza', \\piazzale Aldo Moro 5, 00185 Roma, Italy\\
e-mail: imponente@icra.it, montani@icra.it
\vspace*{4cm}

\vspace{0.5cm}

{\bf   Abstract  \\ } \end{center} \indent
We analyze the dynamics of the Mixmaster Universe on the base of a standard Arnowitt-Deser-Misner Hamiltonian approach showing how its asymptotic evolution to the cosmological singularity is isomorphic to a billiard on the Lobachevsky plane. The key result of our study consists in the temporary gauge invariance of the billiard representation, once provided the use of very general Misner-Chitr\'e-like variables.
 
\end{titlepage}

\section{Introduction}

Since Belinski-Kalatnikov-Lifshitz (BKL) derived the oscillatory regime which characterizes  the behaviour of the Bianchi type VIII and IX cosmological models \cite{BKL70} - \cite{BKL82} (the so-called Mixmaster universe \cite{M69}) near a physical singularity, a wide literature faced over the years this subject in order to provide the best possible understanding of the resulting chaotic dynamics. 

The research activity developed overall in two different, but related, directions: on one hand the dynamical analysis was devoted to remove the limits of the BKL approach due to its discrete nature (by analytical treatments \cite{B82}-\cite{BBT} and numerical simulations \cite{B93}-\cite{B91}), on the other one to get a better characterization of the Mixmaster chaos (especially in view of its properties of covariance \cite{FFM91}-\cite{H94}).

The first line of investigation provided satisfactory representations of the Mixmaster dynamics in terms of continuous variables (leading to the construction of an invariant measure for the system \cite{CB83}, \cite{KM97}), as well as detailed numerical descriptions (allowing to make precise validity tests on the obtained analytical results) \cite{B94}.

The efforts to develop a precise characterization of the chaoticity observed in the Mixmaster dynamics found non-trivial difficulties due to the impossibility, or in the best cases the ambiguity, to apply the standard chaos indicators to relativistic systems. In particular, the puzzle consisting of numerical simulations which were providing, zero \cite{FF92} and non-zero \cite{SK93} Lyapunov exponents by using different time variables, has been solved by realizing the non-covariant nature of these indicators and their inapplicability to hyperbolic manifolds \cite{GK86}. The existence of these difficulties prevented, up to now, to say a definitive word about the covariance of the 
Mixmaster chaos, with particular reference to the possibility of removing the observed chaotic features by a suitable choice of the time variable (apart from the indication provided by \cite{CL97}). 

Indeed a valuable framework of analysis of the Mixmaster evolution, able to join together the two considered points of view, relies on an Hamiltonian treatment of the dynamics in terms of Misner-Chitr\'e-like variables \cite{C72}. This formulation  allows to individualize the existence of an asymptotic (energy-like) constant of motion when performed an ADM reduction. By this result the stochasticity of the Mixmaster can be treated either in terms of the statistical mechanics (by the microcanonical ensemble), either by its characterization as isomorphic to a billiard on the two-dimensional Lobachevsky space \cite{A89}. 

The aim of this work is to show how the representation of the Mixmaster dynamics as a ``stochastic scattering'' isomorphic to a billiard on the Lobachevsky space can be constructed independently of the choice of a time variable, simply providing very general Misner-Chitr\'e-like coordinates. Up to the limit of the adopted approximation on the form of the potential term, our analysis shows, without any ambiguity, that the Mixmaster stochasticity can not be removed by any redefinition of the time variable. 

More precisely in Section $2$ we construct the Hamiltonian formulation of the Mixmaster dynamics, which in Section $3$, by a standard Arnowitt-Deser-Misner (ADM) reduction and the study of the asymptotic form of the potential, allows to individualize an appropriate energy-like constant of motion (nothing more then the ADM kinetic energy) in any temporal gauge. In Section $4$ the reduced Hamiltonian principle is rewritten in terms of a geodesic one on the induced Jacobi metric. Finally in Section $5$ we derive the statistical implications due to the properties of geodesic flow and provide brief concluding remarks. 
      
\section{The Hamiltonian Formulation}

The geometrical structure of the Bianchi type VIII and IX spacetimes, i.e. of the so-called Mixmaster Universe models, is summarized by the line element \cite{BKL70}

\begin{equation} 
\label{a} 
ds^2=-{N(\eta)}^2d{\eta}^2+e^{2\alpha}\left(e^{2\beta}\right)_{ij}\sigma^i \sigma^j 
\end{equation} 
where $N(\eta)$ denotes the lapse function, $\sigma^i$ are the dual 1-forms associated with the anholonomic basis \footnote{ The dual 1-forms of the considered models are given by: \\[1em]

(Bianchi ~VIII): $\left\{ 
\begin{array}{lll} 
\sigma ^1 =-\sinh \psi \sinh\theta d\phi~ + ~\cosh \psi d\theta \\
\sigma ^2= -\cosh \psi \sinh\theta d\phi ~+~\sinh \psi d\theta \\
\sigma ^3=~\cosh\theta d\phi ~+ ~d \psi 
\end{array} \right.$ 
\\[1em]

(Bianchi ~IX): \quad  $\left\{ 
\begin{array}{lll} 
\sigma ^1 = ~\sin \psi \sin\theta d\phi ~+~\cos \psi d\theta \\
\sigma ^2 = -\cos \psi \sin\theta d\phi ~+~\sin \psi d\theta \\
\sigma ^3 = ~\cos\theta d\phi ~+~d \psi  
\end{array} \right.$
}
and $\beta_{ij}$ is a traceless $3\times 3$ symmetric matrix  ${\rm diag}(\beta_{11},\beta_{22},\beta_{33})$; $\alpha$, $N$, $\beta_{ij}$ are functions of $\eta$ only. Parameterizing the matrix $\beta_{ij}$ by the usual Misner variables \cite{M69}
\begin{eqnarray}
\beta_{11}&=&\beta_+ + \sqrt3 \beta_- \nonumber \\
\beta_{22}&=&\beta_+ - \sqrt3 \beta_-   \\
\beta_{33}&=&-2 \beta_+ \nonumber 
\label{a1}
\end{eqnarray} 
the dynamics of the Mixmaster model is described by a canonical variational principle

\begin{equation} 
\label{b} 
\delta I=\delta\int L\, d\eta=0 \, ,
\end{equation} 
with Lagrangian $L$ 

\begin{equation} \label{r} 
L=\frac{6 D}{N}\left[{-{\alpha}^{\prime}}^2+{{\beta_+}^{\prime}}^2+{{\beta_-}^{\prime}}^2\right]- \frac{N}{D}V\left(\alpha, \beta_+ , \beta_-\right) \, .
\end{equation}  
Here ${()}^{\prime} = \frac{d}{d\eta}$, $D\equiv \det e^{\alpha +\beta_{ij}}=e^{3\alpha}$  and the potential $V\left(\alpha, \beta_+ , \beta_-\right)$ reads

\begin{equation} \label{a2} 
V=\frac{1}{2} \left( D^{4H_1}+D^{4H_2}+D^{4H_3}\right) -D^{2H_1 +2H_2}\pm D^{2H_2 +2H_3}\pm D^{2H_3 +2H_1} \, ,
\end{equation}  
where $(+)$ and $(-)$ refers respectively to Bianchi type VIII and IX, and the anisotropy parameters $H_i ~(i=1,2,3)$ denote the functions \cite{KM97}
\begin{eqnarray}
\label{ssaa}
H_1 &=& \frac{1}{3}+ \frac{\beta_+ + \sqrt{3} \beta_-}{3 \alpha} \nonumber \\
H_2 &=& \frac{1}{3}+ \frac{\beta_+ - \sqrt{3} \beta_-}{3 \alpha}   \\
H_3 &=&\frac{1}{3}- \frac{2\beta_+}{3 \alpha}  \nonumber \, .
\end{eqnarray} 
In the limit $D\rightarrow 0$ the second three terms of the above potential turn out to be negligible with respect to the first one.
Let's introduce the new (Misner-Chitr\'e-like) variables 
\begin{eqnarray}
\label{f2}
&\alpha =& -e^{f\left(\tau\right)}\xi \nonumber \\
&\beta_+ =& e^{f\left(\tau\right)}\sqrt{\xi^2 -1}\cos \theta \\
&\beta_- =& e^{f\left(\tau\right)}\sqrt{\xi^2 -1}\sin \theta  \, ,\nonumber 
\end{eqnarray} 
with $f$ denoting a generic functional form of $\tau$, $1\le \xi <\infty$ and $0\le \theta < 2 \pi$. Then the Lagrangian (\ref{r}) reads
\begin{equation} 
L=\frac{6 D}{N} \left[ \frac{{\left(e^f {\xi}^{\prime}\right)}^2}{\xi ^2 -1} +{\left(e^f {\theta}^{\prime}\right)}^2\left(\xi ^2 -1\right) -{{\left(e^f\right)}^{\prime}}^2 \right]  -\frac{N}{D}V \left( f\left(\tau\right), \xi, \theta \right) \, .
\label{g2} 
\end{equation} 
In terms of $f\left(\tau\right)$, $\xi$ and $\theta$ we have
\begin{equation} 
D= exp\left\{ -3 \xi \cdot e^{f\left(\tau\right)} \right\}
\label{h} 
\end{equation} 
and since $D \rightarrow 0$ toward the singularity, independently of its particular form, in this limit $f$ must approach infinity.
The Lagrangian (\ref{r}) leads to the conjugate momenta 
\begin{eqnarray}
p_{\tau}&=&\frac{\partial L}{\partial {\tau}^{\prime}} = -\frac{12 D}{N}{\left(e^f \cdot \frac{df}{d\tau}\right)}^2  {\tau}^{\prime}   \nonumber \\
p_{\xi}&=&\frac{\partial L}{\partial {\xi}^{\prime}} = \frac{12 D}{N} \frac{e^{2f}}{{\xi}^2 -1}{\xi}^{\prime}  \\
p_{\theta}&=&\frac{\partial L}{\partial {\theta}^{\prime}} = \frac{12 D}{N}e^{2f}\left({\xi}^2 -1\right)  {\theta}^{\prime} \nonumber 
\label{i}
\end{eqnarray} 
which by a Legendre transformation make the variational principle (\ref{b}) assume the Hamiltonian form  
\begin{equation} 
\delta \int \left(   p_{\xi} {\xi}^{\prime} +  p_{\theta} {\theta}^{\prime}+    p_{\tau}  {\tau}^{\prime} - \frac{Ne^{-2f}}{24 D} {\cal H}                            \right) d\eta =0 \, ,
\label{m} 
\end{equation} 
being 
\begin{equation} 
{\cal H} = -\frac{{p_{\tau}}^2}{\left(\frac{df}{d\tau}\right)^2} + 
 {p_{\xi}}^2\left(\xi ^2 -1\right) +\frac{{p_{\theta}}^2}{\xi ^2 -1} +24 V e^{2f}  \, .
\label{n} 
\end{equation}

\section{Reduced Variational Principle}

By variating (\ref{m}) with respect to $N$ we get the constraint ${\cal H} =0$, which solved provides 
\begin{equation} 
-p_{\tau}\equiv \frac{df}{d\tau}\cdot {\cal H}_{ADM} = \frac{df}{d\tau} \cdot \sqrt{\varepsilon ^2 +24 V e^{2f}}
\label{n2} 
\end{equation}
where
\begin{equation}
\varepsilon ^2 = \left({\xi}^2 -1\right){p_{\xi}}^2 +\frac{{p_{\theta}}^2}{{\xi}^2 -1} 
\label{d2} 
\end{equation} 
in terms of which the variational principle (\ref{m}) reduces to 
\begin{equation} 
\delta \int \left(   p_{\xi} {\xi}^{\prime} +  p_{\theta} {\theta}^{\prime} - {f}^{\prime}{\cal H}_{ADM} \right) d\eta =0 \, .
\label{q} 
\end{equation} 
Since the equation for the temporal gauge actually reads
\begin{equation} 
N\left(\eta\right)= \frac{12 D}{{\cal H}_{ADM}} e^{2f}  \frac{df}{d\tau}  {\tau}^{\prime} \, ,
\label{rs} 
\end{equation} 
our analysis remains fully independent of the choice of the time variable until the form of $f$ and ${\tau}^{\prime}$ is not fixed.

The variational principle (\ref{q}) provides  the Hamiltonian equations for ${\xi}^{\prime}$ and ${\theta}^{\prime}$ 
\footnote{In this study the corresponding equations for $p^{\prime}_{\xi}$ and $p^{\prime}_{\theta}$ are not relevant.}
\begin{eqnarray}
{\xi}^{\prime}&=& \frac{f^{\prime}}{{\cal H}_{ADM}}\left(\xi ^2 -1\right)p_{\xi}           \nonumber \\
{\theta}^{\prime}&=& \frac{f^{\prime}}{{\cal H}_{ADM}} \frac{p_{\theta}}{\left(\xi ^2 -1\right)} \, .
\label{s}
\end{eqnarray}
Furthermore can be straightforward derived the important relation
\begin{equation} 
\frac{d\left({\cal H}_{ADM}f^{\prime}\right)}{d\eta} = \frac{\partial \left({\cal H}_{ADM}f^{\prime}\right)}{\partial\eta} \Longrightarrow \frac{d\left({\cal H}_{ADM}f^{\prime}\right)}{df} = \frac{\partial \left({\cal H}_{ADM}f^{\prime}\right)}{\partial f} \, ,
\label{t} 
\end{equation} 
i.e. explicitly
\begin{equation} 
\frac{\partial{\cal H}_{ADM}}{\partial f}=
\frac{e^{2f}}{2 {\cal H}_{ADM}} 24\cdot \left( 2V+ \frac{\partial V}{\partial f} \right) \, .
\label{u} 
\end{equation} 
In this reduced Hamiltonian formulation, the functional $f\left(\eta\right)$ plays simply the role of a parametric function of time and actually the anisotropy parameters $H_i$ $(i=1,2,3)$ are functions of the variables $\xi, \theta$ only 

\begin{eqnarray} 
\label{v4}
H_1 &=& \frac{1}{3} - \frac{\sqrt{\xi ^2 - 1}}{3\xi }\left(\cos\theta + \sqrt{3}\sin\theta \right)  \nonumber \\ 
H_2 &=& \frac{1}{3} - \frac{\sqrt{\xi ^2 - 1}}{3\xi }\left(\cos\theta - \sqrt{3}\sin\theta \right)  \\ 
H_3 &=& \frac{1}{3} + 2\frac{\sqrt{\xi ^2 - 1}}{3\xi } \cos\theta \, .  \nonumber 
\end{eqnarray}

Finally, toward the singularity ($D \rightarrow 0$ i.e. $f \rightarrow \infty$) by the expressions (\ref{a2}, \ref{h}, \ref{v4}), we see that \footnote{By $O()$ we mean terms of the same order of the inclosed ones.}
\begin{equation}
\frac{\partial V}{\partial f} = O\left(e^f V\right) \, .
\label{z0}
\end{equation} 
Since in the domain $\Gamma_H$, where all the $H_i$ are simultaneously greater than 0, the potential term $U\equiv e^{2f} V$ can be modeled by the potential walls

\begin{eqnarray}
\label{aa}
U_\infty = 
&\Theta _\infty \left(H_1\left(\xi, \theta\right)\right) + \Theta _\infty \left(H_2\left(\xi, \theta\right)\right) + \Theta _\infty \left(H_3\left(\xi, \theta\right)\right) \, \\ 
 &\Theta _\infty \left(x\right) = \left\{ 
\begin{array}{lll} 
+ \infty & if & x < 0 \\ 
\quad 0 & if & x >  0 
\end{array} \right.\nonumber
\end{eqnarray}
therefore in $\Gamma_H$ the ADM Hamiltonian becomes (asymptotically) an integral of motion
\begin{equation}
\forall \{\xi, \theta\}\in{ \Gamma_H} \quad
\left\{ 
\begin{array}{lll} 
{\cal H}_{ADM}= \sqrt{\varepsilon ^2 +24\cdot U} \cong \varepsilon =E =const. \\ 
\frac{\partial {\cal H}_{ADM}}{\partial f} =\frac{\partial E}{\partial f} =  0  \, .
\end{array}
\right.
\label{bb}
\end{equation}

The key point for the use of the Misner-Chitr\'e-like variables relies on the independence of the time variable for the anisotropy parameters $H_i$. 

\section{The Jacobi Metric and the Billiard Representation}

Since above we have shown that asymptotically to the singularity ($f \rightarrow \infty$, i.e. $\alpha \rightarrow  -\infty$) $d{\cal H}_{ADM}/df =0$ i.e. ${\cal H}_{ADM} =\epsilon =E=const.$, the variational principle (\ref{q}) reduces to
\begin{equation}
\delta \int \left( p_{\xi} d\xi + p_{\theta} d\theta -Edf \right) =\delta \int \left(  p_{\xi} d\xi + p_{\theta} d\theta \right)=0 \, ,
\label{cc}
\end{equation}
where we dropped the third term in the left hand side since it behaves as an exact differential.

By following the standard Jacobi procedure \cite{A89} to reduce our variational principle to a geodesic one, we set ${x^a}^{\prime} \equiv g^{ab}p_b$, and by the Hamiltonian equation (\ref{s}) we obtain the metric
\begin{eqnarray}
g^{\xi \xi} &=&\frac{f^{\prime}}{E}\left({\xi}^2 -1\right) \nonumber \\
g^{\theta \theta} &=&\frac{f^{\prime}}{E} \frac{1}{{\xi}^2 -1} \, .
\label{dd}
\end{eqnarray}
By these and by the fundamental constraint relation
 \begin{equation}
\left({\xi}^2 -1\right){p_{\xi}}^2 +\frac{{p_{\theta}}^2}{{\xi}^2 -1} =E^2 \, ,
\label{ee}
\end{equation}
we get 
\begin{equation}
g_{ab}{x^a}^{\prime} {x^b}^{\prime} =\frac{f^{\prime}}{E} \left\{ \left({\xi}^2 -1\right){p_{\xi}}^2 +\frac{{p_{\theta}}^2}{{\xi}^2 -1}\right\}=f^{\prime}  E \, .
\label{ff}
\end{equation}
By the definition ${x^a}^{ \prime}= \frac{dx^a}{ds} \frac{ds}{d\eta}\equiv u^a \frac{ds}{d\eta}$, the (\ref{ff}) rewrites
\begin{equation}
g_{ab}u^a u^b \left( \frac{ds}{d\eta} \right) ^2  = f^{\prime} E \, ,
\label{gg}
\end{equation}
which leads to the key relation
\begin{equation}
d\eta = \sqrt{ \frac{g_{ab}u^a u^b}{f^{\prime} E }}~ds \, .
\label{hh}
\end{equation}
Indeed the expression (\ref{hh}) together with $p_{\xi} \xi^{\prime} +p_{\theta} \theta^{\prime}=Ef^{\prime}$ allows us to put the variational principle (\ref{cc}) in the geodesic form:
\begin{equation}
\delta \int  f^{\prime} E ~d\eta  = \delta \int  \sqrt{ g_{ab}u^a u^b f^{\prime} E} ~ds= \delta \int  \sqrt{ G_{ab}u^a u^b}~ds =0
\label{ii}
\end{equation}
where the metric $G_{ab} \equiv f^{\prime} E g_{ab}$ satisfies the normalization condition $G_{ab}u^a u^b =1$ and therefore \footnote{We take the positive root since we require that the curvilinear coordinate $s$ increases monotonically with increasing value of $f$, i.e. approaching the initial cosmological singularity.} 
\begin{equation}
\frac{ds}{d\eta}=Ef^{\prime}\Rightarrow \frac{ds}{df} =E \, .
\label{ll}
\end{equation}
Summarizing, in the region $\Gamma_H$ the considered dynamical problem reduces to a geodesic flow on a two dimensional Riemannian manifold described by the line element
\begin{equation}
ds^2 =E^2 \left[ 	\frac{d{\xi }^2}{{\xi}^2 -1}+  \left(\xi^2 -1\right) d {\theta }^2 \right] \, .
\label{mm}
\end{equation}
Now it is easy to check that the above metric has negative curvature, since the associated curvature scalar reads $R=-\frac{2}{E^2}$;
therefore the point-universe moves over a negatively curved bidimensional space on which the potential wall (\ref{a2}) cuts the region $\Gamma_{H}$. By a way completely independent of the temporal gauge we provided a satisfactory representation of the system as isomorphic to a billiard on a Lobachevsky plane \cite{A89}.

\section{Invariant Lyapunov Exponent}

In order to characterize the dynamical instability of the billiard in terms of an invariant treatment (with respect to the choice of the coordinates $\xi$, $\theta$), let us introduce the following (orthonormal) tetradic basis
\begin{eqnarray} 
v^i &=&\left(\frac{\sqrt{{\xi}^2-1}}{E}, 0\right)         \nonumber \\
w^i &=&\left(0,\frac{1}{E \sqrt{{\xi}^2-1}}\right)       \,  .
\label{nn}
\end{eqnarray} 
Indeed the vector $v^i$ is nothing else than the geodesic field, i.e. 
\begin{equation}
\frac{Dv^i}{ds}=\frac{dv^i}{ds}+\Gamma^i_{kl}v^k v^l =0 \, ,
\label{oo}
\end{equation}
while the vector $w^i$ is parallely transported along the geodesic, according to the equation
\begin{equation}
\frac{Dw^i}{ds}=\frac{dw^i}{ds}+\Gamma^i_{kl}v^k w^l =0 \, ,
\label{pp}
\end{equation}
where by $\Gamma^i_{kl}$ we denoted the Christoffel symbols constructed by the metric (\ref{mm}).
Projecting the geodesic deviation equation along the vector $w^i$ (its component along the geodesic field $v^i$ does not provide any physical information about the system instability), the corresponding connecting vector (tetradic) component $Z$ satisfies the following equivalent equation 
\begin{equation}
\frac{d^2 Z}{ds^2}=\frac{Z}{E^2} \, .
\label{qq}
\end{equation}
This expression, as a projection on the tetradic basis, is a scalar one and therefore completely independent of the choice of the variables.
Its general solution reads
\begin{equation}
Z\left(s\right)=c_1 e^{\frac{s}{E}}+c_2 e^{-\frac{s}{E}} \, , \qquad c_{1,2}=const. \,  ,
\label{rr}
\end{equation}
and the invariant Lyapunov exponent defined as \cite{PE77} 
\begin{equation}
\lambda_v =\sup \lim_{s\rightarrow \infty} \frac{\ln\left(Z^2+ \left(\frac{dZ}{ds}\right)^2\right)}{2s} \, ,
\label{ss}
\end{equation}
in terms of the form (\ref{rr}) takes the value
\begin{equation}
\lambda_v =\frac{1}{E} > 0 \, .
\label{tt}
\end{equation}
When the point-universe bounces against the potential walls, it is reflected from a geodesic to another one thus making each of them unstable. Though up to the limit of our potential wall approximation, this result shows without any ambiguity that, independently of the choice of the temporal gauge, the Mixmaster dynamics is isomorphic to a well described chaotic system. Equivalently, in terms of the BKL representation, the free geodesic motion corresponds to the evolution during a Kasner epoch and the bounces against the potential walls to the transition between two of them. By itself, the positive Lyapunov number (\ref{tt}) is not enough to ensure the system chaoticity, since its derivation remains valid for any Bianchi type model; the crucial point is that for the Mixmaster (type VIII and IX) the potential walls reduce the configuration space to a compact region ($\Gamma_H$), ensuring that the positivity of $\lambda_v$ implies a real chaotic behaviour (i.e. the geodesic motion fills the entire configuration space).

\vspace{0.8cm}

Summarizing, our analysis shows that for any choice of the time variable, we are able to give the above stochastic representation of the Mixmaster model, provided the factorized coordinate transformation in the configuration space
\begin{eqnarray}
&\alpha &= -e^{f\left(\tau\right)} a\left(\theta , \xi\right) \nonumber\\
&\beta_+ &=~e^{f\left(\tau\right)} b_+ \left(\theta , \xi\right) \\
&\beta_- &=~e^{f\left(\tau\right)} b_- \left(\theta , \xi\right) \, , \nonumber 
\label{uu}
\end{eqnarray}
where $f,a,b_{\pm}$ denote generic functional forms of the variables $\tau, \theta, \xi$.

It is worth noting that the success of our analysis, in showing the time gauge independence of the Mixmaster chaos, relies on the use of a standard ADM reduction of the variational principle (which reduces the system by one degree of freedom) and overall because, adopting Misner-Chitr\'e-like variables, the asymptotic potential walls are fixed in time. The difference between our approach and the one presented in \cite{SL90} (see also for a critical analysis \cite{BT93}) consists effectively in these features, though in those works is even faced the problem of the Mixmaster chaos covariance with respect to the choice of generic configuration variables. 

\vspace{0.8cm}
We are very grateful to Remo Ruffini for his valuable comments on this subject.

\end{document}